\documentclass[graybox, nosecnum]{svmult} 

\usepackage{mathptmx}       
\usepackage{helvet}         
\usepackage{courier}        
\usepackage{type1cm}        
\usepackage{makeidx}         
\usepackage{graphicx}        
\usepackage{multicol}        
\usepackage{multirow}
\usepackage[bottom]{footmisc}
\usepackage[hidelinks]{hyperref}        
\usepackage{soul}            
\usepackage[sort&compress]{natbib}
\usepackage{amsmath,amssymb,mathtools}
\usepackage{doi}
\usepackage{xspace}
\usepackage{aas_macros}

\newcommand{\neff}{\ensuremath{N_{\rm eff}}\xspace}
\newcommand{\gstar}{\ensuremath{g_{\star}}\xspace}
\newcommand{\gstars}{\ensuremath{g_{\star S}}\xspace}

\makeindex             

\begin{document}
\title*{Big Bang Nucleosynthesis}
\author{Evan Grohs \thanks{corresponding author} and George M. Fuller}
\institute{Evan Grohs \at Department of Physics, North Carolina State University, Raleigh, NC 27695, \email{ebgrohs@ncsu.edu}
\and George M. Fuller \at Department of Physics, University of California San Diego, La Jolla, CA, 92093, \email{gfuller@ucsd.edu}}
%
%
\maketitle
\abstract{
 As the early universe expands and cools the rates of the weak interactions that keep neutrinos in thermal equilibrium with the matter and the related rates of the reactions that inter-convert neutrons and protons decrease. Eventually, these rates fall below the expansion rate \--- they freeze out. Likewise, the rates of the strong and electromagnetic nuclear reactions that build up and tear down nuclei, though fast enough to maintain equilibrium early on, slow down and ultimately lead to freeze out. Together these freeze out processes comprise the epoch of Big Bang Nucleosynthesis (BBN). The relics emerging from this early time include the light element abundances, for example of helium and deuterium, and a background of decoupled neutrinos, a \lq\lq C$\nu$B\rq\rq , roughly analogous to the Cosmic Microwave Background, the CMB. These fossil relics encode the history of the physics operating in the early universe. Consequently, BBN has emerged as a key tool for constraining new, beyond-standard-model (BSM) physics. BBN may become an even finer probe of BSM physics, given the anticipated higher precision in measurements of the primordial abundances of deuterium and helium afforded by the advent of large optical telescopes and Stage-4 CMB experiments. The latter experiments will also provide higher precision determinations of $N_{\rm eff}$, a measure of the relativistic energy density at the photon decoupling epoch and, hence, an important probe of the C$\nu$B. }       


\section{Introduction}

The success of Big Bang Nucleosynthesis (BBN) theory in predicting the primordial abundances of helium and deuterium and the baryon (ordinary matter) content of the universe represents one of the greatest triumphs of modern physics [see \cite{1977ARNPS..27...37S,Tytler:2000qf,2000PhR...333..389O,2000A&AT...19..367S,2009PhR...472....1I,2010ARNPS..60..539P,osti_1209735,2014arXiv1412.1408F,Cyburt:2015mya} for reviews of the various physical phenomena present in BBN]. It is all the more remarkable that this success is born of very simplistic assumptions about the universe and its evolution. These are: (1) General Relativity (GR) is a correct description of spacetime dynamics and that the distribution of mass-energy on any 3-dimensional spacelike hypersurface at a given value of the time $t$ (age of the universe) is homogeneous and isotropic; and (2) that the standard model of particle physics and, more specifically, simple nuclear physics obtains at very early times in the history of the universe. 

In fact, the Friedmann-LeMa\^itre-Robertson-Walker metric, the solution to the field equations in the symmetry implied by homogeneity and isotropy [see \cite{1973grav.book.....M}], was worked out in \cite{1917SPAW.......142E} shortly after Einstein's original work on GR. \cite{1922ZPhy...10..377F} and \cite{1924ZPhy...21..326F} showed that the solutions to the GR field equations led to a non-static universe.  It was \cite{1927ASSB...47...49L} and \cite{1939Natur.143..116G} who first took this solution seriously as a description of the history of the universe, and \cite{1929PNAS...15..168H} who first showed astronomical evidence of an expanding universe. In particular, they realized that this solution implied a hot and high energy density early phase in the evolution of the universe. Not only did this realization lead them to predict a decoupled relic radiation component (effectively the cosmic microwave background, the CMB), but it also led to speculation on nuclear reactions and the nuclear abundances that might emerge from an early epoch where the temperature could be on the energy scale of nuclear physics, $T \sim 1\,{\rm MeV}$.

\cite{1949PhRv...75.1089A}, \cite{1949RvMP...21..367G}, \cite{1950PThPh...5..224H}, and others outlined the basic picture of nuclear reaction freeze out and nucleosynthesis in the early universe.  \cite{1966PhRvL..16..410P} and \cite{1966ApJ...146..542P} worked out key issues in helium synthesis. The role of weak interactions and the details of how the nucleosynthesis of the light elements depended on the baryon-to-photon ratio \--- the key input parameter in standard BBN \--- were worked out by \cite{1967ApJ...148....3W} and \cite{1969ApJS...18..247W}.
  
Since then, BBN has been honed into a foundational tool in cosmology [see the textbook treatments in \cite{Kolb:1990vq} and \cite{2020moco.book.....D}; and \cite{1998RvMP...70..303S} for a review]. This was done through increasingly sophisticated determinations of primordial light element abundances and by advances in the experimental nuclear and weak interaction reaction physics input to BBN calculations.  In particular, \cite{1984ApJ...281..493Y} used observations of primordial abundances to infer the baryon to photon ratio based on detailed BBN calculations.
  
At the heart of the success of this enterprise lies an assertion that the large scale geometry of the universe is characterized by a simple symmetry. That assumed symmetry of homogeneity and isotropy in the distribution of matter and radiation at any time leads to two results. 
  
First, this symmetry leads directly to the Friedmann equation. This equation is tantamount to the total mechanical energy (kinetic plus \lq\lq gravitation potential energy\rq\rq) being a constant of the motion for a co-moving 2-spherical surface. The Friedmann equation gives the time rate of change of the scale factor in terms of the mass-energy density and the spatial curvature parameter (itself related to a scaled total mechanical energy of a co-moving two sphere).
\begin{equation}
  {\dot{a}}^2+ k = {{8 \pi}\over{3 }}\, G\, \rho\, a^2
  \label{Friedmann}
\end{equation}
where the scale factor is $a$, ${\dot{a}} = da/dt$, the gravitational constant is $G=1/m_{\rm pl}^2$ with $m_{\rm pl}\simeq1.2\times10^{19}\,{\rm GeV}$ the Planck mass (here we use natural units), and the mass-energy density is $\rho$.
Observations of the expansion history of the universe, e.g., scale factor as a function of redshift or time, can then allow the Friedmann equation to be \lq\lq reverse engineered\rq\rq\ to give a history of the mass-energy density and the total energy on any co-moving 2-sphere (i.e., determining whether that 2-sphere is gravitationally bound or will expand on forever). The best fit to the observations is with that total energy being zero, corresponding to curvature parameter $k=0$.  

Second, demanding that the symmetry of homogeneity and isotropy always obtain means that there can be no preferred spacelike directions. This, in turn, means that there can be no spacelike heat flow [a timelike, uniform source or sink of heat, however, is still consistent with this symmetry; see \cite{1971ApJ...168..175W}]. For example, there cannot be a current of energy or heat in some spacelike direction. If there is no heat flow of {\it any kind} then the evolution of the early universe is adiabatic, meaning the entropy in a co-moving 2-sphere is constant. In turn, that implies a simple relationship between the temperature of the matter and radiation and the scale factor. If $S$ is the total proper entropy density, then constant total entropy on co-moving 2-spheres corresponds to the condition that $S\, a^3$ is constant. If the temperature is low enough that baryon number can be regarded as conserved, then $S\, a^3$ being constant is equivalent to the entropy-per-baryon $s=S/n_b$ being constant. Here $n_b$ is the proper baryon number density.
  
The baryon-to-photon ratio $\eta \equiv n_b/n_\gamma$ can be ascertained through the CMB anisotropies (i.e., through ratio of the amplitudes of the acoustic peaks in the CMB power spectrum) to be $\eta \approx 6.1\times{10}^{-10}$ as inferred by the \cite{PlanckVI_2018}. This determination is consistent with the independent BBN-derived value based on the observationally-inferred primordial deuterium abundance in \cite{2003ApJS..149....1K,2014ApJ...781...31C,2016ApJ...830..148C,2018ApJ...855..102C}, to be discussed later. Armed with this number, we can conclude that in the standard model the conditions in the BBN epoch, broadly defined, will be radiation dominated. That is, the mass-energy density and entropy will be carried by particles with relativistic kinematics. In this limit, we write $\rho = (\pi^2/30)\, \gstar\, T^4$ and $S = (2 \pi^2/45)\, \gstars\, T^3$. Here \gstar is the energetic statistical weight in relativistic particles, given by
\begin{equation}\label{def_g}
  \gstar = \sum_b{g_b}\left(\frac{T_b}{T}\right)^4+{{7}\over{8}} \sum_f{g_f}\left(\frac{T_f}{T}\right)^4,
\end{equation}
while \gstars is the entropic statistical weight in relativistic particles, given by
\begin{equation}\label{def_gs}
  \gstars = \sum_b{g_b}\left(\frac{T_b}{T}\right)^3+{{7}\over{8}} \sum_f{g_f}\left(\frac{T_f}{T}\right)^3.
\end{equation}
In both Eqs.\ \eqref{def_g} and \eqref{def_gs} the sums are over Bose-Einstein ($b$) and Fermi-Dirac ($f$) degrees of freedom, $T_b$ is the temperature-like quantity for a given bosonic species, and $T_f$ for fermionic [see Eqs.\ (3.62) and (3.73) in \cite{Kolb:1990vq}]. For most of the history of the early universe, $\gstar=\gstars$.  A notable exception is post-neutrino decoupling when the temperature is $T\lesssim 1\,{\rm MeV}$
\begin{align}
  \gstar&=2+{{7}\over{8}} \left( 2+2+1+1+1+1+1+1\right) = 10.75, \label{g_today}\\
  \gstars&=2+{{7}\over{8}} \left( 2+2\right) = 5.5, \label{gs_today}
\end{align}
For $T\sim1\,{\rm MeV}$ the photon ($g_b=2$) bath is accompanied by relativistic distributions of electrons ($g_f =2$), positrons ($g_f =2$) and, for this example, assumed equal numbers (zero chemical potentials) for neutrino species $\nu_e$,$\bar\nu_e$,$\nu_\mu$,$\bar\nu_\mu$,$\nu_\tau$,$\bar\nu_\tau$, each contributing $g_f=1$ to \gstar.  Opposed to this, \gstars receives the same contributions from photons, electrons, and positrons, but no such contribution from any neutrino species. In both Eqs.\ \eqref{g_today} and \eqref{gs_today}, overall electric charge neutrality implies that the electron chemical potential is small, of order the baryon-to-photon ratio. There is no such constraint on the net lepton numbers (electron, muon, tau) in Eq.\ \eqref{g_today}, as these can reside in asymmetries between the number densities in the seas of neutrinos and antineutrinos. In fact, our best constraints on these asymmetries come from the observationally-inferred primordial abundance of helium or deuterium together with standard homogeneous BBN models [see \cite{2000NuPhB.590..539E,2002NuPhB.632..363D,2004NJPh....6..117K,2017PhRvD..95f3503G}].

The Friedmann equation, Eq.\ \eqref{Friedmann}, and the entropy condition can be solved together to give the time, temperature, scale factor history of the early universe {\it if} we know the mass spectrum and decay and interaction properties of particles and the dynamics of the vacuum. If we know how to calculate $\rho$, \gstar, $s$, and \gstars given the temperature, then these two relations summarize the expansion and thermal histories:
\begin{equation}\label{hub_g}
  H={{\dot{a}}\over{a}}=\left({{8 \pi}\over{3 }}\right)^{1/2} {{ \rho^{1/2}}\over{m_{\rm pl} }}\approx \left({{8 \pi^3}\over{90 }}\right)^{1/2}\, g_{\star}^{1/2}\, {{T^2}\over{m_{\rm pl}}}\  \ \ {\rm and}\ \ \ a\, T\, \propto {{s^{1/3}}\over {g_{\star S}^{1/3} }}    
\end{equation}
where the approximation in the first equation and the form of the second expression assume a radiation dominated energy density and entropy. Here $H$ is the Hubble expansion rate. 

In the standard model, with no Beyond-Standard-Model (BSM) physics, and no vacuum dynamics (e.g., cosmic vacuum phase transitions, inflation, etc.), these approximations will be valid at the $T \sim 1\,{\rm MeV}$ energy scale of BBN. Note that the product of scale factor and temperature is fixed for a regime in temperature where $s$ and \gstar do not change. However, it must be kept in mind that timelike heat flows from out-of-equilibrium decay of BSM particles or out-of-equilibrium scattering of particles can change the entropy $s$. The latter process does indeed occur for standard model neutrinos, though the magnitude of the associated entropy change is small [see \cite{1997NuPhB.503..426D}, \cite{2016PhRvD..93h3522G} and \cite{2018PhR...754....1P}]. In any case, an effective boundary condition on the simultaneous solution of the expressions in Eq.\ \eqref{hub_g} is that an evolutionary history in the early universe must hit the $\eta$-inferred entropy-per-baryon $s \approx 5.9\times{10}^9$ (in units of Boltzmann's constant $k_{\rm b}$ per baryon) by the time of photon decoupling at $T=T_\gamma^{\rm dec} \approx 0.2\,{\rm eV}$. 

The Hubble expansion rate is relatively slow. This is because gravitation is weak and it sets the scale for the expansion rate, as is obvious from Eq.\ \eqref{Friedmann}. That slow expansion provides plenty of time for very weakly-interacting particles, like neutrinos, to come into equilibrium and contribute to dynamics. Moreover, the entropy is high (on a nuclear physics scale), meaning there are large numbers of photons, electrons, positrons, and neutrinos per baryon. Together slow expansion and high entropy team up to enable the BBN epoch and its relic observables to comprise a \lq\lq laboratory\rq\rq\ for probing and constraining new physics. The history of the neutrino component provides a concrete example.

\section{Neutrino and Weak Interaction Decoupling}

The neutrinos are weakly interacting particles that can come into thermal and chemical equilibrium in the early universe.
In equilibrium at high temperature, $T > {\cal{O}}({\rm MeV})$, neutrino rest masses are negligible and so their number densities will be similar to those of photons and are crudely $\sim T^3$. The expansion rate from Eq.\ \eqref{Friedmann} in these conditions will scale as $H \sim g_{\star}^{1/2} T^2/m_{\rm pl}$. By contrast, the weak interaction charged and neutral current neutrino scattering, absorption and emission rates are $\sim G_{\rm F}^2 T^5$. The Fermi constant, $G_{\rm F}\approx 1.166\times{10}^{-11}\,{\rm MeV}^{-2}$, sets the scale for weak interaction rates. The different temperature dependence of the expansion rate and the weak interaction rates means that as the universe expands and the temperature drops, at some point the neutrino interaction rates will fall below the expansion rate, and the neutrino component will cease to interact in the age of the universe $\sim 1/H$ \---- meaning that the neutrinos are decoupled and in free fall through spacetime. 

Thermal neutrino decoupling, where the neutrinos cease to scatter rapidly enough to exchange energy efficiently with the photon-electron/positron-baryon plasma, and chemical decoupling, where the lepton capture-induced neutron-proton inter-conversion rates fall well below $H$, proceed over relatively lengthy time scales. These decoupling epochs, sometimes termed Weak Decoupling and Weak Freeze-Out, respectively, actually occur roughly concurrently. They are lengthy in the sense that they play out over hundreds of Hubble times $H^{-1}$, between $T\sim 10\,{\rm MeV}$ and $T\sim 0.1\,{\rm MeV}$. 

Neutrino and antineutrino scattering on electrons and positrons is the principal channel for energy exchange between the neutrino component and the electron, positron, nucleon, photon plasma. The entropy-per-baryon is high enough that the number of electron-positron pairs in electromagnetic equilibrium will be larger than the number of ionization electrons (i.e., those required for charge neutrality) down to temperatures $T \sim 20\,{\rm keV}$, which is more than an order of magnitude below the $2 m_e$ threshold for the radiation field to make a pair [see \cite{2020PhRvD.101f3507T}]. Nevertheless, at high entropy there are plenty of photons on the tail of the Planck distribution that have energies above this threshold even when $T \ll 2 m_e$. Eventually and inevitably the weak interaction $\nu-{e^\pm}$ scattering rate will fall below the Hubble rate for $T < 1\,{\rm MeV}$. This is Weak Decoupling.

\begin{figure}
\begin{center}
\includegraphics[width=12cm]{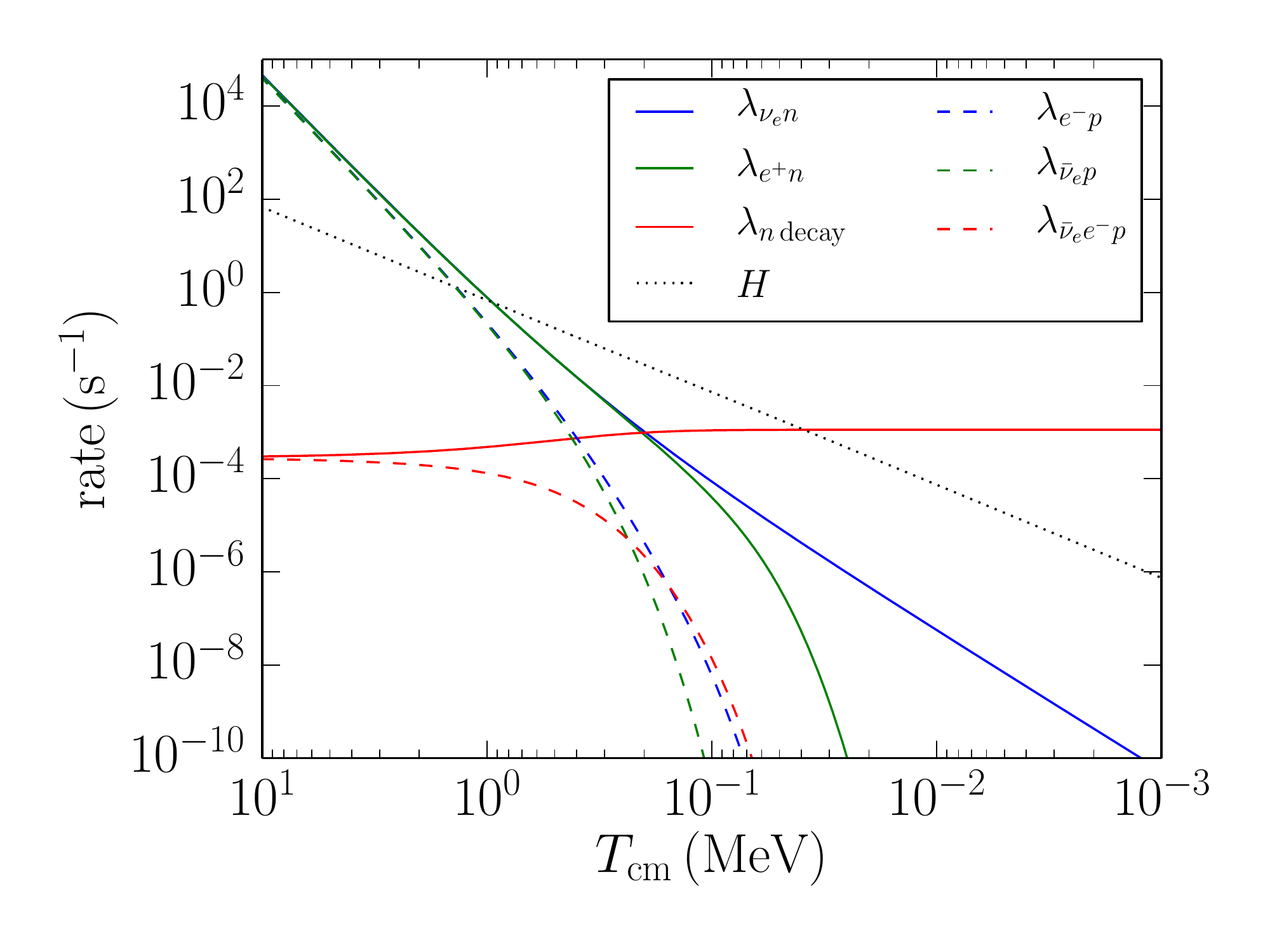}
\end{center}
\caption{ Rates for lepton capture and decay processes that inter-convert protons and neutrons [Eqs.\ \eqref{np1} -- \eqref{ndecay}] as a function of co-moving temperature $T_{\rm cm}$ (a proxy for inverse scale factor, $1/a$). The Hubble expansion rate $H$ is also shown [Eq.\ \eqref{hub_g}]. Legend provides a key for the different curves. Figure from 
\cite{2016NuPhB.911..955G}.
}
\label{fig1:nprates}
\end{figure}

The isospin-changing charged current weak interactions involving the leptons and nucleons are:
\begin{subequations}
\begin{align}
&\nu_e + n \rightleftharpoons p + e^- \label{np1}\\
&\bar\nu_e + p \rightleftharpoons n + e^+ \\
&n \rightleftharpoons p + e^- + \bar\nu_e. \label{ndecay}
\end{align}
\end{subequations}
These reactions do not contribute significantly to energy transfer between the neutrino component and the plasma because the baryon-to-photon ratio $\eta$ is so small. However, the neutrino-to-baryon ratio is of order $\sim 1/\eta$, so there are more than a billion neutrinos of each kind per nucleon. At high temperature the rates of the two-body lepton capture reactions will be fast compared to $H$, so chemical equilibrium will obtain. Put another way, at high temperature the rate at which the isopspin of a nucleon flips between neutron $n$ and proton $p$ is fast compared to $H$. This flipping rate eventually slows below $H$. That is Weak Freeze-Out. The neutron-proton mass difference is $\delta m_{np} \approx 1.29\,{\rm MeV}$. This mass difference represents a threshold in the charged current channel that converts a proton to a neutron. Consequently, at lower temperatures the rates of electron capture and $\bar\nu_e$ capture on protons, $\lambda_{e p}$ and  $\lambda_{\bar\nu_e p}$, respectively, fall below the corresponding neutron destruction rates, $\lambda_{\nu_e n}$ and $\lambda_{e^+ n}$. Even after Weak Freeze-Out the neutron-to-proton ratio will slowly decrease because of free neutron decay, i.e., $\lambda_{n\,{\rm decay}}$, the forward process in Eq.\ \eqref{ndecay}, augmented by neutrino capture [see \cite{2016NuPhB.911..955G}], the process in Eq.\ \eqref{np1}. Note that the rate for this decay process is always slower than that for free vacuum neutron decay because of the Pauli blocking effects of the $\bar\nu_e$ sea and, to a lesser extent at lower $T$, the $e^-$ sea. The rates for all of these processes and the Hubble expansion rate are shown as functions of $T_{\rm cm}$ (a proxy for inverse scale factor and close to the temperature $T$) in Fig.\ \ref{fig1:nprates}.

We can follow the rates of the forward and reverse reactions in Eqs.\ \eqref{np1} -- \eqref{ndecay} and solve
\begin{equation}
{{d }\over{ dt}}\left(n/p\right) = n_p\,{\left( \lambda_{e^- p}+\lambda_{\bar\nu_e p} +\lambda_{\bar\nu_e p e^-}\right)} - n_n\,{\left( \lambda_{e^+ n}+\lambda_{\nu_e n} +\lambda_{n\, {\rm decay}} \right)}
\label{n/p}
\end{equation}
to find the neutron-to-proton number density ratio  $n/p \equiv n_n/n_p$ as a function of time $t$. In chemical equilibrium $n/p$ will be unity for temperatures well in excess of the neutron-proton mass difference, but will fall as the temperature drops and it becomes more energetically favorable for baryon number to reside in protons rather than in heavier neutrons. In the higher temperature, $T > 0.7\, {\rm MeV}$, equilibrium regime $n/p = \exp{\{(-\delta m_{n p} -\mu_{\nu_e} +\mu_e  )/T\}}$, where the electron chemical potential is of order $\eta$ and is insignificant, but where the electron neutrino chemical potential $\mu_{\nu_e}$ could play a role [see \cite{2002PhR...370..333D}]. If we take $\mu_{\nu_e}=0$ (zero net electron lepton number residing in the electron flavor neutrino and antineutrino seas), then $n/p \approx 1/6$ at temperatures $T \approx 0.7\,{\rm MeV}$ and then will slowly fall to $\approx 1/7$ by the time $T \approx 0.1\,{\rm MeV}$.

\section{Nuclear Freeze Out}


The history of the baryonic component is dictated in part by this weak interaction-driven evolution of isospin, e.g., as embodied in the $n/p$ ratio. However, the interplay of entropy and the rates of strong and electromagnetic nuclear reactions largely sets the scale for how many nucleons assemble into nuclei during BBN. The substantial Coulomb barriers for charged particle nuclear reactions imply that the rates of these reactions are sensitive functions of temperature: nuclear reactions go faster at high temperature. At temperatures in excess of $T \sim 0.1\,{\rm MeV}$ the rates of nuclear reactions that build up and tear down nuclei are balanced in equilibrium, and both are fast compared to the Hubble expansion $H$. This is Nuclear Statistical Equilibrium (NSE). As the universe expands and cools, the nuclear reactions slow down and, eventually, are no longer able to maintain NSE. In a sense, primordial nucleosynthesis is a freeze out from NSE. This point bears particular emphasis as freeze out from equilibrium is an irreversible process.  As a result, the asymptotic abundances are sensitive to the time-evolutions of $n/p$ and the entropy, in addition to the initial abundances at the point of departure from NSE.

The abundance relative to baryons $Y_A$ of a nucleus with mass number $A$ and binding energy $B_E$ in NSE at entropy-per-baryon $s$ (in units of Boltzmann's constant $k_{\rm b}$) and temperature $T$ is given by a Saha equation as alluded to in \cite{1957RvMP...29..547B}. The nuclear Saha equation relates the total chemical potentials (including mass) of free neutrons $\mu_n$, free protons $\mu_p$, with the chemical potential $\mu_A$ for a nucleus with $A=Z+N$, corresponding to the reaction and concomitant equilibrium equation
\begin{subequations}
\begin{align}
&Z p + N n \rightleftharpoons A +\gamma,
\label{reactions}\\
\implies &Z \mu_p+N \mu_n = \mu_A
\label{saha}
\end{align}
\end{subequations}
In practice, the schematic expression in Eq.\ \eqref{reactions} and equilibrium equation in Eq.\ \eqref{saha} summarizes a potentially extensive network of reactions that build up and destroy nucleus $A(Z,N)$.

\begin{figure}
\begin{center}
\includegraphics[width=12cm]{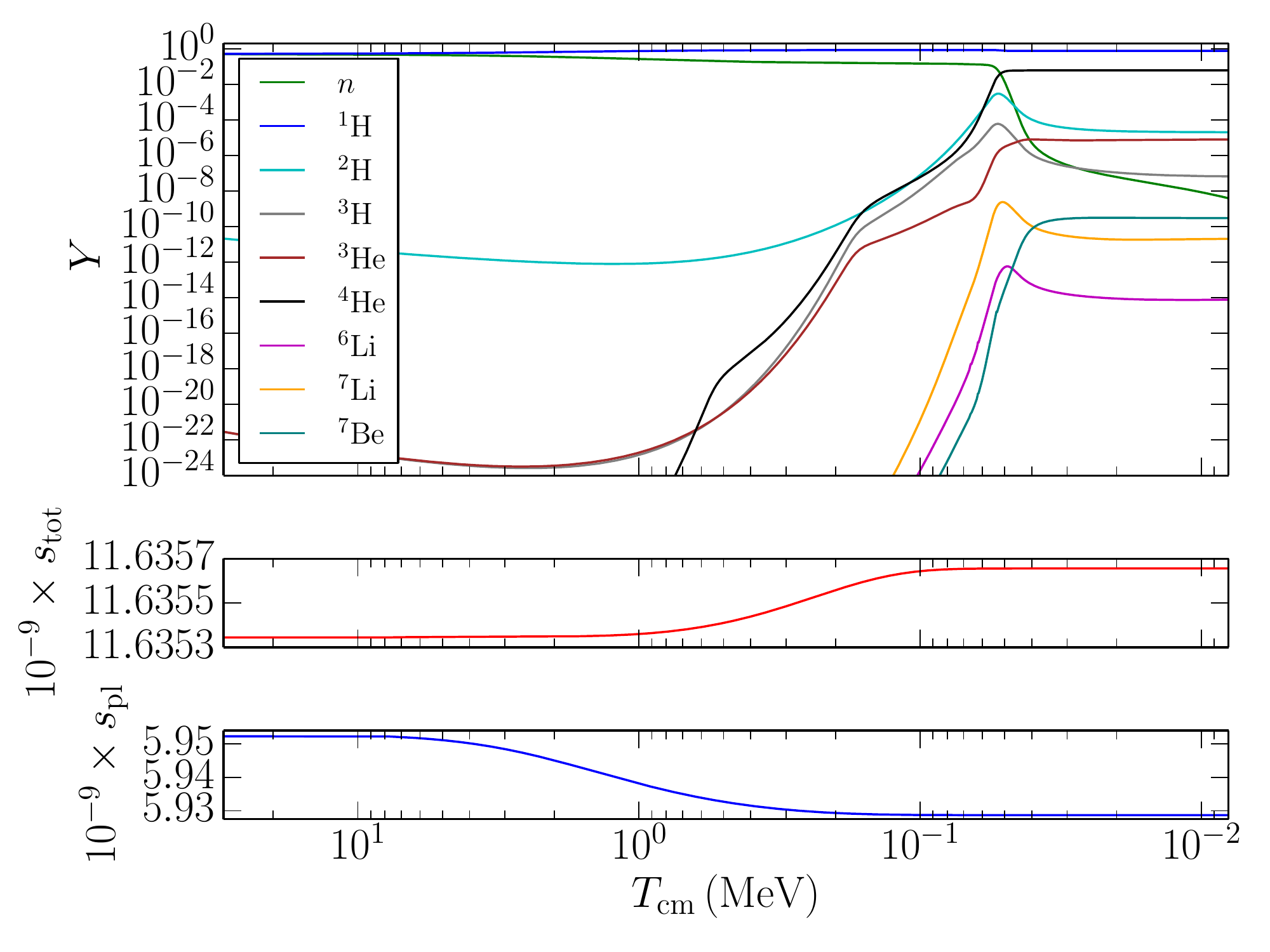}
\end{center}
\caption{BBN abundances $Y$ (see legend) for light nuclear species, total entropy per baryon (plasma plus the decoupling neutrino component) $s_{\rm tot}$, and entropy in the photon-electron/positron-baryon plasma $s_{\rm pl}$, are each shown as a function of co-moving temperature $T_{\rm cm}$ (a proxy for inverse scale factor, $1/a$) in the top, middle, and lower panels, respectively. Figure from \cite{2016PhRvD..93h3522G}.
}
\label{fig14:abund}
\end{figure}

Omitting the dependence on nuclear charge, $n/p$, and nuclear partition functions on the chemical potentials in Eq.\ \eqref{saha}, we can give a heuristic version of this Saha equation, one that starkly illustrates the fight between binding and disorder that characterizes NSE in the early universe
\begin{equation}
Y_A \propto s^{1-A}\, \exp{(B_E/T)}.\label{saha2}
\end{equation}
Were the freeze out from NSE instantaneous at a temperature $T_{\rm fo}$, Eq.\ \eqref{saha2} would predict the BBN abundance yields. However, in line with the physics of the weak interaction during Weak Decoupling and Weak Freeze-Out, Nuclear Freeze-Out is not abrupt and different individual reactions will freeze out at different times, right in the regime where the neutrino and charged lepton components are driving an evolution of $n/p$. An accurate prediction of the BBN abundance yields then demands a simultaneous and self-consistent calculation of all relevant strong, electromagnetic and weak nuclear reactions together with a sufficiently accurate treatment of weak interactions and the neutrino component. This was first done in \cite{1967ApJ...148....3W}, updated in \cite{1973ApJ...179..343W}, and further updated in \cite{1993ApJS...85..219S}. Results of a calculation with a more sophisticated treatment of the weak interaction physics which include a Boltzmann transport scheme for neutrino scattering-mediated energy and entropy flow are shown in Fig.\ \ref{fig14:abund} from \cite{2016PhRvD..93h3522G}.
See \cite{Pisanti:2007hk,2018PhR...754....1P,2020CoPhC.24806982A,2022CoPhC.27108205G} for public BBN codes.

The key features of BBN abundance yields can be gleaned from Eq.\ \eqref{saha2} and our discussion of $n/p$ evolution. At the high entropy of the standard model early universe, the fight in NSE between disorder and binding is won decisively by alpha particles. Not surprisingly, this NSE prediction carries over in broad brush to detailed nuclear reaction network calculations. These indeed show that nearly every nucleon that can be incorporated into an alpha particle {\it is} locked away there. This assembly of alphas takes place rather abruptly when the temperature falls to $T\approx 0.08\,{\rm MeV}$ for $\eta = 6.1\times{10}^{-10}$.
The asymptotic value of the helium mass fraction can be calculated using the freeze-out value of $n/p$, assuming all neutrons end up incorporated in alpha particles
\begin{equation}\label{yp}
  Y_{\rm P}\equiv 4Y_{^4{\rm He}}\vert_{\rm fo} = \frac{2n_n}{n_p + n_n} = \frac{2n/p}{1+n/p}
\end{equation}
For a freeze-out value of $n/p=1/7$, $Y_{\rm P}\simeq25\%$.
Not all primordial neutrons reside in $^4{\rm He}$ nuclei.  However, the alpha particle assembly process is very efficient at scouring out the neutrons, with only about one neutron in $\sim{10}^5$ incorporated into a deuteron, and with one in $\sim {10}^{9}$ ending up in a $^7{\rm Be}$ or $^7{\rm Li}$ nucleus, justifying the approximation in Eq.\ \eqref{yp}. 

The sense of the evolution of the deuteron ($d$, or ${^2{\rm H}}$ in Fig.\ \ref{fig14:abund}) abundance throughout the NSE freeze out epoch can also be understood simply from these considerations. In equilibrium, the Saha equation corresponding to
\begin{equation}\label{npg}
 p + n \rightleftharpoons d +\gamma,
\end{equation}
predicts that the deuteron abundance rises exponentially as the temperature falls, $Y_d \propto \exp{(2.2\,{\rm MeV}/T)}$ where $B_E \approx 2.2\,{\rm MeV}$ for the deuteron (see the cyan curve in Fig.\ \ref{fig14:abund} for $1.0 > T_{\rm cm}/{\rm MeV}>0.06$). This reaction is rather abruptly taken out of equilibrium when the neutrons are incorporated into alpha particles. This deprives the forward, deuteron assembly reaction in Eq.\ \eqref{npg} of neutron \lq\lq fuel,\rq\rq\ leading to unbalanced, but quite modest destruction of deuterons predominantly through the electromagnetic and strong reactions
\begin{subequations}
\begin{align}
  p + d &\rightleftharpoons {}^3{\rm He} + \gamma,\\
  d + d &\rightleftharpoons {}^3{\rm He} + n,\\
  d + d &\rightleftharpoons {}^3{\rm H} + p,
\end{align}
\end{subequations}
and to a lesser extent the reverse of Eq.\ \eqref{npg}, namely $d(\gamma,p)n$ as evidenced in \cite{2014PhRvD..90b3543D}.
Deuterium and the $A=3$ nuclides are further depleted by the nuclear reactions to synthesize $^4{\rm He}$
\begin{subequations}
\begin{align}
  d + {^3{\rm He}} &\rightleftharpoons {{}^4{\rm He}} + p\\
  d + {^3{\rm H}}&\rightleftharpoons {{}^4{\rm He}} + n.
\end{align}
\end{subequations}
The overall result is that a higher baryon density (higher $\eta$) leads to {\it earlier} (higher $T$) assembly of alphas, and hence the Saha equation prediction of a {\it lower} deuterium yield, and {\it vice versa} for a lower baryon density (lower $\eta$).
$d(d,n){^3{\rm He}}$ and its isospin mirror $d(d,p){^3{\rm H}}$, and the subsequent chain of reactions in the alpha particle assembly network, e.g., ${^3{\rm He}}(d,p)\alpha$ and its mirror ${^3{\rm H}}(d,n)\alpha$. The overall result is that a higher baryon density (higher $\eta$) leads to {\it earlier} (higher $T$) assembly of alphas, and hence the Saha equation prediction of a {\it lower} deuterium yield, and {\it vice versa} for a lower baryon density (lower $\eta$).



Overall, the BBN abundance yield predictions from detailed reaction network calculations, e.g., as shown in Fig.\ \ref{fig14:abund}, for helium and deuterium are in good agreement with observationally-inferred values of the primordial abundances of these species. High redshift damped Lyman-alpha system hydrogen absorption lines were used to infer the primordial deuterium abundance in \cite{1998ApJ...499..699B,1998ApJ...507..732B,1999ApJ...519...18B,2014ApJ...781...31C,2016ApJ...830..148C,2018ApJ...855..102C}. This constituted the first measurement (or inference) of $\eta$ in \cite{1996Natur.381..207T}, \cite{2003ApJS..149....1K}. Later, this measurement was {\it confirmed} by the observations of the CMB anisotropy  acoustic peak amplitude ratios in \cite{2003ApJS..148....1B}.
For $^3{\rm He}$, no measurement exists at high redshift to compare with theoretical predictions.  A measurement of the {\it galactic} $^3{\rm He}$ abundance in \cite{2002Natur.415...54B} can be used to set an upper limit on the cosmological abundance.  The upper bound is consistent with the primordial abundances of deuterium and helium-4 with a CMB-inferred value of $\eta$.  In addition, \cite{2022ApJ...932...60C} has used galactic chemical-evolution modeling to evolve the ratio of $^3{\rm He}/^4{\rm He}$.  Using the current value of the isotopic ratio in the Orion Nebula, \cite{2022ApJ...932...60C} reverse solves for the initial assumed primordial ratio. The chemical-evolution results agree within $2\sigma$ of standard BBN predictions.
Observations of primordial Helium-4 come from ionized hydrogen regions in small, hot galaxies.  These regions contain helium and are metal-deficient.  $Y_{\rm P}$ is deduced by inferring what the $^4{\rm He}$ abundance is at zero metalicity -- the ostensible value for the metalicity in BBN
[see \cite{2013JCAP...11..017A}, \cite{2014MNRAS.445..778I}].

The Spite plateau in the lithium {\it versus} surface temperature curves obtained in observations of old (lower metallicity) halo stars, from \cite{1982A&A...115..357S}, suggests a primordial $^7{\rm Li}$ abundance that is a factor of $3$ or $4$ below the BBN-predicted value for the $^7{\rm Be}+^7{\rm Li}$ yield (beryllium decays to lithium shortly before the photon decoupling epoch).
When evaluating lithium as a function of metalicity (using iron content), \cite{2010A&A...522A..26S} also observed the Spite plateau.
Resolution of this \lq\lq 
lithium problem\rq\rq\, as detailed in \cite{2011ARNPS..61...47F}, may lie in the interpretation of observations and in stellar physics, or in beyond standard model (BSM) physics, but it is unlikely to be found in standard nuclear physics [see \cite{PhysRevD.74.085008} and \cite{PhysRevD.82.105005}].
Most lithium from BBN resides in the $A=7$ isotope.  $^6{\rm Li}$ is synthesized at a level less than 1 part in $10^4$ compared to $^7{\rm Li}$ [see the magenta line in Fig.\ \ref{fig14:abund}].  \cite{2006ApJ...644..229A} claimed a detection of the ratio $^6{\rm Li}/^7{\rm Li}$ of a few percent -- in strong tension with the predicted BBN ratio.  Later,  \cite{2007A&A...473L..37C} noted that non-local thermodynamic equilibrium effects in the modeling of stellar atmospheres are important when identifying the blending of the $^6{\rm Li}$ with the $^7{\rm Li}$ line, implying a non-detection of the isotope shift.  \cite{2013A&A...554A..96L} came to the same conclusion as \cite{2007A&A...473L..37C} when including additional isotopes in their stellar atmospheric models.  In any case, the result of \cite{2006ApJ...644..229A} should only be taken as an upper limit for $^6{\rm Li}/^7{\rm Li}$ which is consistent with the prediction from BBN.

Nuclear reaction and weak interaction physics that is important as input for BBN calculations is being addressed in current and future laboratory experiments, such as the LUNA experiment in \cite{2020Natur.587..210M}. For example, LUNA's very low background site and accelerator/detector set-up allow higher precision low energy nuclear reaction cross section measurements. A case in point is LUNA's recent measurement of $d(p,\gamma){^3{\rm He}}$
cross section at BBN energies.
The new nuclear reaction data has reaffirmed concordance between CMB and BBN [see \cite{2021JCAP...04..020P} and \cite{2021JCAP...03..046Y}], however, small tensions may exist when a global analysis of $d(p,\gamma)^3{\rm He}$ reaction data is used [see \cite{2021MNRAS.502.2474P}].  Future data on the transfer reactions $d(d,p)^3{\rm H}$ and $d(d,n)^3{\rm He}$ could shed light on these tensions [see \cite{2021NatRP...3..231P}].

As discussed above, the physics of the charged current weak interaction is foundational for BBN. The vector coupling in lepton-nucleon weak interactions is well measured. However, the axial vector weak coupling, which is obtained from the neutron lifetime, remains a front line target for research. The advent of ultra cold neutron sources has transformed this experimental effort, allowing ultra high precision measurements. The high precision in magnetic \lq\lq bottle\rq\rq\ experiments such as \cite{2021PhRvL.127p2501G}, where neutrons are counted, allows tests of unitarity in the Cabibbo-Kobayashi-Maskawa (CKM) quark mixing matrix. These measurements suggest a $3\sigma$ discrepancy from unity in the \lq\lq first row\rq\rq\ CKM absolute square-matrix element sum, $\vert V_{\rm u d}\vert^2 +\vert V_{\rm u s}\vert^2 +\vert V_{\rm u b}\vert^2$, potentially signaling BSM physics. Also potentially significant for revealing BSM physics, the bottle experiment results are discrepant with the \lq\lq beam\rq\rq\ experiments of \cite{2013PhRvL.111v2501Y}, where protons from beta decay are counted rather than neutrons. This discrepancy has, for example, been interpreted as a small dark sector neutron decay branch [see \cite{2018PhRvL.120s1801F} for the original model, and see \cite{2022PhRvD.105k5005A} for a discussion of constraints on, and ramifications of, this model].

\section{Entropy, the Evolution of the Neutrino Component, and the Cosmic Neutrino Background (C$\nu$B)}

The history of entropy through the weak decoupling and NSE freeze out epochs is foundational for BBN and is key to understanding the features of the relic neutrino background, the C$\nu$B [see \cite{2016PhRvD..93h3522G} and \cite{2018PhR...754....1P}] and other cosmological environments [see \cite{1971ApJ...168..175W}]. The relation on the right in Eq.\ \eqref{hub_g}, valid whenever the bulk of the entropy is carried by particles with relativistic kinematics, shows how the product of scale factor and temperature evolves in terms of the time or temperature history of the entropy-per-baryon, $s$, and the entropic statistical weight in relativistic particles, \gstars in Eq.\ \eqref{hub_g}.

First consider the case where there is no timelike heat flow and so $s$ is fixed. Moreover specify that the plasma of the early universe is populated only by standard model particles. As the temperature drops below the masses of these particles, they will cease to contribute to \gstars, and so the product $a T$ will increase. In the BBN epoch this scenario does indeed play out, with electrons and positrons dropping out of significant contribution to \gstars when $T\ll 2 m_e$. The entropy carried by $e^\pm$-pairs at high temperature is transferred to the photons, but not to any decoupled particles, like neutrinos. 

Decoupled neutrinos by definition have ceased scattering and are simply free-falling through spacetime, with their 3-momenta, $p$, redshifting like $p\propto 1/a$. These free-falling particles then remain described by a (relativistic) Fermi-Dirac black body-shaped energy distribution but with \lq\lq temperature\rq\rq\ at scale factor $a$ given by $T_\nu = T_\nu^{\rm DEC}\,(a^{\rm DEC}/a)$, where the temperature at which they decouple (last scattered) is $T_\nu^{\rm DEC}$, corresponding to scale factor $a^{\rm DEC}$. Assuming that neutrino decoupling occurs when $e^\pm$-pairs in electromagnetic equilibrium are large in number, and are relativistic, then
\begin{equation}
  g_{\star S}^{\rm (DEC)}=2+\frac{7}{8}(2+2)= \frac{11}{2},
\end{equation}
as in Eq.\ \eqref{gs_today}. Of course, the photon-electron-positron temperature is the same as the neutrino temperature prior to neutrino decoupling. At low temperature, after the $e^\pm$-pair numbers have been suppressed by their masses, photons carry all of the entropy and $g_{\star S}^{\rm (low)}=g_\gamma=2$. If the co-moving entropy $s$ is constant, then Eq.\ \eqref{hub_g} shows that the ratio of the neutrino temperature to the plasma temperature is
\begin{equation}
\left(\frac{g_{\star S}^{\rm (low)}}{g_{\star S}^{\rm (DEC)}}\right)^{1/3} = \left(\frac{2}{11/2}\right)^{1/3} = \left(\frac{4}{11}\right)^{1/3} \approx 0.714.
\end{equation}

This simple picture predicts a relic decoupled neutrino background for each neutrino species, with a temperature about $40\%$ lower than the CMB photon temperature. Degeneracy parameters (the ratio of chemical potential to temperature), $\eta_{\nu_\alpha}$, with $\alpha=e,\mu,\tau$, are co-moving invariants. The 3-momentum distribution functions for the relic neutrinos are then characterized by a temperature and a degeneracy parameter
\begin{equation}
f_{\nu_\alpha} \approx {{1 }\over{T^3_{\nu_\alpha} F_2(\eta_{\nu_\alpha})} }\cdot {{p^2}\over{e^{(p/T_{\nu_\alpha}-\eta_{\nu_\alpha}) }+1 }}
\label{dist}
\end{equation}
where $F_2$ is a relativistic Fermi integral of order 2 and argument $\eta_{\nu_\alpha}$, so that $\int_0^\infty f_{\nu_\alpha}\, dp=1$ and the local proper number density of $\nu_{\alpha}$'s is $n_{\nu_\alpha} = (T_{\nu_\alpha}^3/2 \pi^2)\,F_2(\eta_{\nu_\alpha})$. 
Observationally-inferred primordial deuterium and helium abundances constrain the neutrino degeneracy parameters to be small, roughly $\eta_{\nu_\alpha} < 0.1$ [see \cite{2004NJPh....6..117K}, \cite{PhysRevD.74.085008}, \cite{2010JCAP...05..037S}].

This standard picture with a constant entropy $s$ is simple and in good agreement with current bounds, but we cannot preclude a timelike source of heat, i.e., a changing $s$ in the early universe. This is obviously the case for BSM scenarios where new particles decay out of equilibrium, for example inflaton decay in inflationary models. 
 
But even in the standard model, out-of-equilibrium scattering of neutrinos provides a timelike heat source, albeit a small one. During the extended BBN epoch, the photon-electron/positron-baryon component has a slightly higher temperature than the neutrino component. Neutrino scattering on electrons and positrons then effects entropy transfer from the plasma to the decoupling neutrino component. Neutrino Boltzmann energy transport calculations [see, e.g., \cite{2016PhRvD..93h3522G} and \cite{2018PhR...754....1P}] demonstrate that about three parts in a thousand of the entropy in the plasma, $s_{\rm pl}$, is transferred to the neutrinos. This entropy transfer is accompanied by an even smaller overall increase of a few parts in ${10}^5$ in the overall {\it total} entropy. This evolution in entropy and its distribution among the neutrino and plasma components is depicted in the lower two panels of Fig.\ \ref{fig14:abund}. This out of equilibrium entropy transfer accompanies correspondingly small distortions in the energy and momenta spectra of the relic neutrinos in the C$\nu$B. This is why the momentum distribution functions in Eq.\ \eqref{dist} are only approximately Fermi-Dirac black bodies.

We do not detect the C$\nu$B directly, but broad agreement between calculated BBN light element abundance yields and observation suggest that it was at least there during NSE freeze out. Moreover, CMB measurements, sensitive to the relative mix of energy density from relativistic and non-relativistic components at the photon decoupling epoch, $T_\gamma^{\rm DEC} \approx 0.2\,{\rm eV}$, also indirectly detect this neutrino component. The energy resident in the relativistic component, $\rho_{\rm rel}$, at this epoch is parameterized by $N_{\rm eff}$
\begin{equation}
\rho_{\rm rel} = \left[ 2+ {{ 7}\over{4 }}\left({{ 4}\over{11 }} \right)^{4/3}\,N_{\rm eff} \right]\,{{ \pi^2}\over{30}}\,(T_\gamma^{\rm DEC})^4
\label{Neff}
\end{equation}
Any relativistic energy source will contribute to $N_{\rm eff}$. If only photons and the standard picture C$\nu$B with zero degeneracy parameters and Fermi-Dirac black body momentum spectra contribute, then we expect $N_{\rm eff} = 3$.
Two standard-model effects act to perturb \neff away from the integer value.  First, finite-temperature QED effects in the electromagnetic plasma [see \cite{1994PhRvD..49..611H}, \cite{1997PhRvD..56.5123F}] change the equation of state for the photon and charged lepton components.  These effects were first estimated in \cite{1982PhRvD..26.2694D} and \cite{1982NuPhB.209..372C} in the context of corrections to the neutron-to-proton rates of Eqs.\ \eqref{np1} -- \eqref{ndecay} and the subsequent impact on the primordial helium abundance [see also \cite{1999PhRvD..59j3502L} for an updated approach to the $n\rightleftharpoons p$ rates].
\cite{2017NuPhB.923..222G} and \cite{2020JCAP...03..003B} examined the same physics but in the context of energy density to arrive at a change in \neff of $\sim0.01$.
The second standard-model effect is due to out-of-equilibrium neutrino scattering-induced spectral distortions arising during weak decoupling.  Early mentions of the effects from energy transport date to \cite{1982PhRvD..26.2694D}, and were followed by \cite{1992PhRvD..46.5378D}, \cite{1993PhRvD..47.4309F}, \cite{1997NuPhB.503..426D} among others.
After \cite{1997NuPhB.503..426D}, the neutrino decoupling problem was revisited by multiple groups, including \cite{2002PhLB..534....8M}, \cite{2015NuPhB.890..481B}, \cite{2016PhRvD..93h3522G} and \cite{2018PhR...754....1P} using Boltzmann neutrino energy transport and ignoring effects from oscillations.  These groups all found similar results, namely, an increase in \neff of $\sim0.034$.
The first work to include neutrino-flavor oscillations in a 3-flavor generalized density matrix formalism, along with the QED corrections to the plasma equation of state, was \cite{2005NuPhB.729..221M}, which found $\neff=3.046$.
More recent calculations which include all standard-model effects (including oscillations) coalesce on
$N_{\rm eff} = 3.044$ [see \cite{2016JCAP...07..051D}, \cite{Akita:2020szl}, \cite{Froustey:2020mcq}, and \cite{Bennett:2020zkv}].
So far this is consistent with CMB bounds in \cite{PlanckVI_2018}, as are the helium and deuterium abundances resulting from BBN calculations using the CMB-derived value of $\eta$. Anticipated large optical telescopes promise sub-one percent precision on the primordial deuterium abundance [see Fig.\ (7) in \cite{2016ApJ...830..148C}]. This will complement results from Stage-4 CMB experiments that are projected to give comparable constraints on $N_{\rm eff}$ and primordial helium, as detailed in \cite{2016arXiv161002743A}.

\section{BBN and the C$\nu$B as \lq\lq Laboratories\rq\rq\ for BSM Physics}

Neutrino (weak interaction) physics and entropy considerations lie at the heart of BBN. In turn, this suggests that BSM physics modifications of standard model physics could have a significant impact. This whets our appetites for a new era of BBN constraints on, and probes of, the physics operating in the early universe [see \cite{2005APh....23..313C,2010ARNPS..60..539P,2016PhRvD..93h3522G,astro2020,2020JCAP...09..051G,2022arXiv220906854G,bond_inprep}].

An important result from the BBN calculations that incorporate neutrino scattering and energy transport is that {\it any} physics that alters the time-temperature-scale factor relation relative to the standard model picture could result in concomitant alterations in light element abundance yields, $N_{\rm eff}$, and the relic energy-momentum spectrum of the C$\nu$B and, hence, $\sum m_\nu$ -- the \lq\lq sum of the light neutrino masses.\rq\rq\. As discussed above, standard model neutrino out-of-equilibrium scattering effects on these quantities are small. This may not be the case for BSM physics, where modifications of the standard model result could range from negligible to dramatic. For a particular BSM model the observable quantities might exceed observational bounds, in which case that model would be constrained. It is conceivable that a BSM model would \lq\lq move\rq\rq\ some calculated quantities relative to their standard model values in a way characteristic to that model. That BSM model could then either be constrained or its signatures searched for [see \cite{astro2020} and \cite{bond_inprep}]. The physics of the neutrino component illuminates the possibilities and promise.

For example, CMB and large scale structure considerations allow probes of $\sum m_\nu$. That quantity encodes the neutrino energy spectrum and the neutrino rest masses [see \cite{2017PhLB..775..239G} and \cite{2019BAAS...51c..64D} for more details on $\sum m_\nu$ in cosmology].  In fact, $\sum m_\nu$ gives a measure of the neutrino collision-less damping scale. This is a gauge of how far neutrinos freely stream and, hence, their effectiveness at damping the growth of smaller-scale structure. \cite{2016arXiv161002743A} forecast $15\,{\rm meV}$ $1\,\sigma$ sensitivity to $\sum m_\nu$. Assuming perfect Fermi-Dirac black body momenta spectra, zero degeneracy parameters, the normal neutrino mass hierarchy, and further assuming the lightest neutrino mass eigenvalue is $m_1=0$, plus adopting the measured neutrino mass-squared differences, would give $\sum m_\nu \approx 57\,{\rm meV}$. At this level we would not expect a conventional neutrino rest mass-mediated spin flip signal in a tonne-scale neutrinoless double beta decay detector, but a signal could arise from BSM physics. However, if $m_1=10\,{\rm meV}$ or greater we might expect a conventional neutrino mass-mediated positive signal, even with the normal neutrino mass hierarchy. This case gives $\sum m_\nu \approx 75\,{\rm meV}$, differing from the $m_1=0$ case by roughly one $\sigma$ in the projected Stage-4 CMB reach. Would the CMB experiments show this value of $\sum m_\nu$? If they did not, then we could ask if BSM lepton number-violating physics is facilitating the neutrinoless double beta decay, or whether BSM physics modifies the nuclear matrix element for this process, or whether there is another astrophysical issue involved in the growth of the smaller scales in the large scale structure of the universe -- or maybe all of these. Or, more likely, the experiments will not reach the sensitivity to definitely establish a problem. This is, obviously, a complicated picture. It is nevertheless tantalizing and, if nothing else, illustrative of the promise of the combined power of next generation CMB experiments, long baseline neutrino oscillation experiments (pinning down the neutrino mass hierarchy), neutrino rest mass experiments, and 30-meter class telescopes [see also \cite{2022arXiv220307377A}].

An obvious possibility for extension of the standard model in the neutrino sector is the introduction of sterile neutrinos [see \cite{2021PhR...928....1D}]. Heavy sterile neutrinos are frequently invoked in neutrino mass models. See-Saw models engineer the very light masses that neutrinos are known to have by positing, for example, that the product of the active neutrino mass and the sterile neutrino mass is a very large mass-squared scale, and then the sterile state mass is taken to be very large. However, these ultra heavy sterile states may not be the only sterile neutrinos. It is conceivable that sterile neutrinos could have much lower masses, in ranges where they would be produced during the BBN epoch if their vacuum mixing with active neutrino states were large enough. These have been suggested as explanations for various experimental anomalies. However, the existence of light, $\sim {\rm eV}$ mass scale sterile neutrinos could be very problematic for BBN if the sterile state is fully thermalized with the active neutrino bath [see in particular the early seminal work by \cite{1977PhLB...66..202S}]. That, in turn, either allows constraints on the mass and mixing properties of these sterile states, or invites speculation on BSM extensions that could suppress the production of these sterile neutrinos in the early universe [see \cite{PhysRevD.74.085008}]. 

Sterile neutrinos with $\sim {\rm keV}$ masses and very small (e.g., $\sim {10}^{-10}$) vacuum mixing with active neutrino have been suggested as a component of dark matter. These could be produced for example, by active neutrino scattering-induced de-coherence in the very early universe (where $T\sim 1\,{\rm GeV}$) [see \cite{1994PhRvL..72...17D,2001PhRvD..64b3501A,2002APh....16..339D,2003PhRvD..68j3002F,2005PhLB..631..151A,2008PhRvD..78b3524K,2017JCAP...01..025A,2017PhR...711....1A,2019PrPNP.104....1B}]. These would have little or no influence on the physics of the BBN epoch as their energy density contribution at the BBN epoch is negligible. However, these dark matter candidates may have tiny admixtures with active neutrino states and that allows for a radiative decay channel. That, in turn, enables X-ray astronomy to provide the best probes and constraints on this speculative sector of particle physics [see \cite{2001ApJ...562..593A}].  For details on X-ray signals with possible sterile neutrino interpretations, see \cite{2014ApJ...789...13B} and \cite{2014PhRvL.113y1301B}.

However, heavier sterile states with small mixing might be created in the early universe by a variety of means. They would decouple early on. If these particles decay out-of-equilibrium during the extended weak decoupling/BBN epoch ($10\,{\rm MeV} \gtrsim T \gtrsim 100\,{\rm keV}$) then they would add entropy in this time frame and so modify the time-temperature-scale factor relationship. That, in turn, could alter $N_{\rm eff}$, the light element abundance yields, and the relic C$\nu$B energy spectrum in ways that allow constraint [see \cite{2011arXiv1110.6479F}, \cite{2020JCAP...09..051G}, \cite{2022PhRvD.105h3513R}, etc.]. 

The future holds promise of improved laboratory measurements of key BBN reaction cross sections, high precision CMB observations, and high precision determinations of primordial deuterium. We believe that this will enable BBN science to be a key way to vet and probe BSM physics.

\section*{Acknowledgements}

G.M.F. acknowledges National Science Foundation (NSF) Grant
No. PHY-2209578 at University of California San Diego and the NSF Network for Neutrinos Nuclear Astophysics and Symmetries (N3AS) Physics
Frontier Center, NSF Grant No. PHY-2020275, and
the Heising-Simons Foundation (2017-228).

\bibliographystyle{apsrev4-2}
\bibliography{references.bib}

\end{document}